# A Pole-to-Pole Pressure-Temperature Map of Saturn's Thermosphere from Cassini Grand Finale Data

Z. Brown[1*], T. Koskinen[1], I. Müller-Wodarg[2], R. West[3], A. Jouchoux[4], L. Esposito[4]

[1]Lunar and Planetary Laboratory, University of Arizona, USA.
[2]Space & Atmospheric Physics Group, Imperial College London, UK.
[3]Jet Propulsion Laboratory, California Institute of Technology, USA.
[4]Laboratory for Atmospheric and Space Physics, University of Colorado, USA.
*e-mail: zbrown@lpl.arizona.edu

**Temperatures of the outer planet thermospheres exceed those predicted by solar heating alone by several hundred degrees. Enough energy is deposited at auroral regions to heat the entire thermosphere but models predict that equatorward distribution is inhibited by strong Coriolis forces and ion drag[1,2]. A better understanding of auroral energy deposition and circulation are critical to solving this so-called energy crisis. Stellar occultations observed by the UVIS instrument during the Cassini Grand Finale were designed to map the thermosphere from pole to pole. We analyze these observations, together with earlier observations from 2016 and 2017, to create a two-dimensional map of densities and temperatures in Saturn's thermosphere as a function of latitude and depth. The observed temperatures at auroral latitudes are cooler and peak at higher altitudes and lower latitudes than predicted by models, leading to a shallower meridional pressure gradient. Under modified geostrophy[3], we infer slower westward zonal winds that extend to lower latitudes than predicted, supporting equatorward flow from approximately 70° to 30° latitude in both hemispheres. We also show evidence of atmospheric waves in the data that can contribute to equatorward redistribution of energy through zonal drag.**

Cassini observed the Grand Finale suite of more than 30 stellar occultations in the UVIS far ultraviolet (FUV) and extreme ultraviolet (EUV) channels between June 24th and August 4th, 2017. These observations probe different longitudes and local times at latitudes from 86° S to 86° N, including over a dozen at polar latitudes. We analyze 23 EUV occultations with sufficient data quality to allow for accurate $H_2$ density and temperature retrieval along with 10 previously unpublished observations from 2016 and earlier in 2017 (Fig. 1a, Supplementary Table 1). These data probe pressures from ~0.1 µbar to ~1 picobar, corresponding to equatorial altitudes of ~1,000 to ~3,000 km above the 1 bar level (see Methods).

We fit $H_2$ density profiles above the altitude of 1,700-1,800 km to retrieve exospheric temperatures as a function of latitude (Fig. 1b). From ~30˚ to ~65˚ latitude, temperature increases from 394 to 513 K in the south and from 340 K to 586 K in the north, consistent with previously published results from Cassini[4] and Voyager[5] occultations. At polar latitudes, the temperature decreases with latitude from 513 K to 373 K between 61˚ S and 86˚ S, and from 586 K to 437 K between 66˚ N and 86˚ N. The Voyager data point at 82˚ S was previously considered an outlier[5] but now agrees well with the observed polar temperature minima. Our temperatures also agree roughly with the temperatures inferred from column-integrated $H_3^+$ observations[6,7]. The magnitude of temporal and longitudinal variations in temperature is limited by the standard deviation, which we infer for 15° latitude bins over the course of the mission. This standard deviation ranges from 11 to 56 K, with the largest value at 30-45˚ N, reflecting a greater temporal variability at those latitudes (Supplementary Figure 1).

We compare the observed temperatures to those predicted by the Saturn Thermosphere-Ionosphere general circulation model[1,2] (Fig. 1b), the only current model of Saturn's thermosphere to include both global circulation and the response to auroral and solar heating.

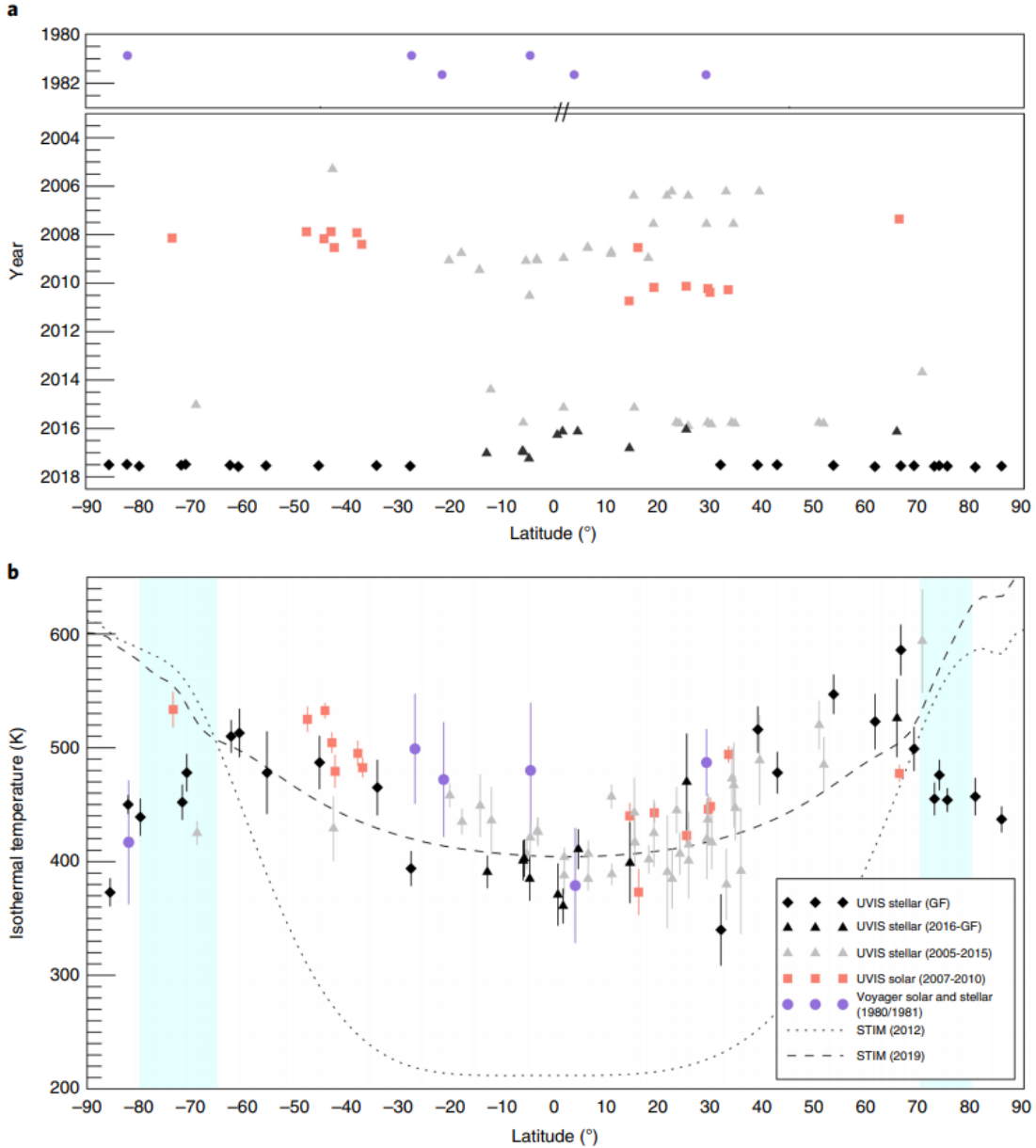

**Fig. 1 | Distribution of exospheric temperatures with latitude. a**, Coverage map of occultation data. Black symbols are from this work: diamonds show UVIS stellar occultation data from the Grand Finale between June 24th and August 4th, 2017 and triangles are previous UVIS stellar occultations from February 14th, 2016 to March 28th, 2017. Grey triangles show previously published UVIS stellar occultation data taken between 2005 and 2015[4,20]. Red squares show UVIS solar occultation data taken between May 6th, 2007 and September 15th, 2010[20]. Purple circles show Voyager data taken between November 3rd 1980 and August 19th, 1981 and analyzed in 2015[5]. **b**, Exospheric temperatures as a function of planetocentric latitude at the half-light point for the observations described above. The dotted and dashed lines show the exospheric temperature profile predicted for northern solstice by the STIM general circulation model, 2012 and 2019 versions, respectively[1,2]. Exospheric temperatures peak at the equatorward edge of the main aural oval (light blue shaded regions)[22]. Error bars on UVIS temperature data represent the 1-σ uncertainty based on the fitting algorithms.

The most recent version of the model (hereafter, the STIM 2019 model) applies zonal drag in the momentum equation to slow down westward jets at high latitudes and allow for the redistribution of auroral energy to match the observed low-latitude temperatures. At polar latitudes, the STIM 2019 model predicts that temperature further increases with latitude or remains flat. While the cool polar temperatures differ from model predictions, similar temperatures observed in the north and south agree with the seasonal trend predicted by the STIM 2019 model for northern summer solstice (May 24, 2017). Additional heating in the southern hemisphere from the northward shift of Saturn's magnetic field[8] offsets the larger Joule heating derived from greater photo-ionization in the northern summer hemisphere, an effect which has also been observed in $H_3^+$ temperatures[9].

The polar temperature minima are difficult to explain in the context of existing models. Joule heating due to auroral currents is expected to be the main heating mechanism at high latitudes. The magnetospheric electric field that drives the predicted auroral currents, however, is constrained to latitudes of ~65 - 90° by ionosphere-magnetosphere interactions and falls to zero inside the polar cap. Therefore, dynamics is predicted to redistribute energy from the auroral oval towards the poles, leading to the higher than observed temperatures inside the polar caps. In terms of cooling mechanisms, radiative cooling is unlikely to play a role in the polar energy balance. This is because we probe altitudes above the homopause, where the abundance of species heavier than $H_2$ that could contribute significantly to radiative cooling is low. Also, $H_3^+$ cooling that is important on Jupiter[12,13,14] is expected to be negligible on Saturn[2]. This means that dynamics, in combination with the auroral electric fields, is still the likely driver of the observed polar temperatures. The zonal winds that we infer from the data extend to lower latitudes than predicted and as a result, the poleward meridional winds are also shifted to lower latitudes (see below). This implies that poleward transport is less efficient than expected. Adiabatic cooling due to polar upwelling would also help to cool the poles, although the data are not sufficient to confirm this. Our results call for a revision of the existing models to properly explain the polar temperature structure, particularly the shift of the observed temperature peak from the predicted location.

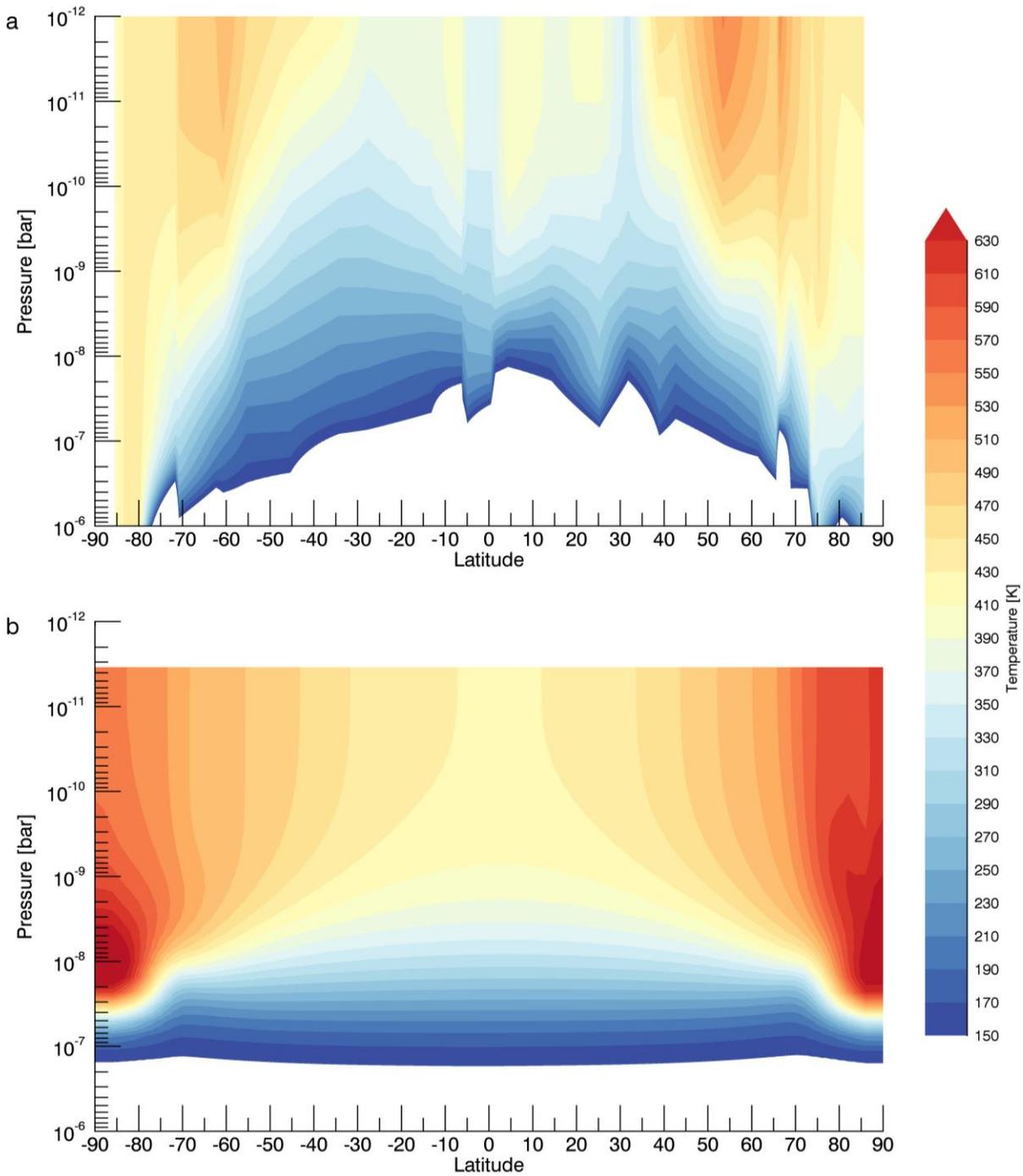

**Fig. 2 | Meridional profile of observed and model temperatures. a**, Forward model temperatures from the 33 occultations analyzed in this work, interpolated over pressure and latitude. The irregular shape of the bottom of the plot follows the lower altitude limit of the retrieval region for each occultation as described in the Methods **b**, STIM 2019 temperature predictions for northern solstice conditions[2] ( northern solstice occurred May 24[th], 2017). The temperature scale at right applies to both plots, with the coldest temperatures at the deepest pressures in both cases.

By interpolating over the observed temperature profiles, we create a latitude-pressure contour (Fig. 2a) and compare the results with the STIM 2019 model (Fig. 2b). The observed temperatures at low latitudes are similar to model predictions (~150-420 K), however, the data differs from the model at higher latitudes. The highest temperatures in the model are located at the poles at a pressure of about $10^{-8}$ bar, with maximum values higher than 650 K. The observations show a cooler maximum temperature (~550 K), located higher in the atmosphere and at lower latitudes than the STIM 2019 model predicts, resulting in shallower meridional pressure gradients. This has implications for winds.

We use the data to infer winds at three pressure levels (10, 1 and 0.1 nbars) under modified geostrophy in oblate spheroidal coordinates[3,15], assuming a balance between ion drag, pressure gradient, and Coriolis forces[3]. The wind equations are populated with our data-derived geopotential gradients along with electric field and conductivity values identical to those used by STIM 2019, allowing for a direct comparison of the resulting wind fields (see Methods). The inferred zonal winds are close to geostrophic and depend mostly on the gravitational potential gradient on constant pressure levels (Fig. 3a). We infer the radii and latitudes required to calculate the potentials from the observed density and temperature profiles. By fitting a polynomial to the potentials, we determine the potential gradient with latitude and calculate the zonal and meridional winds (see Methods).

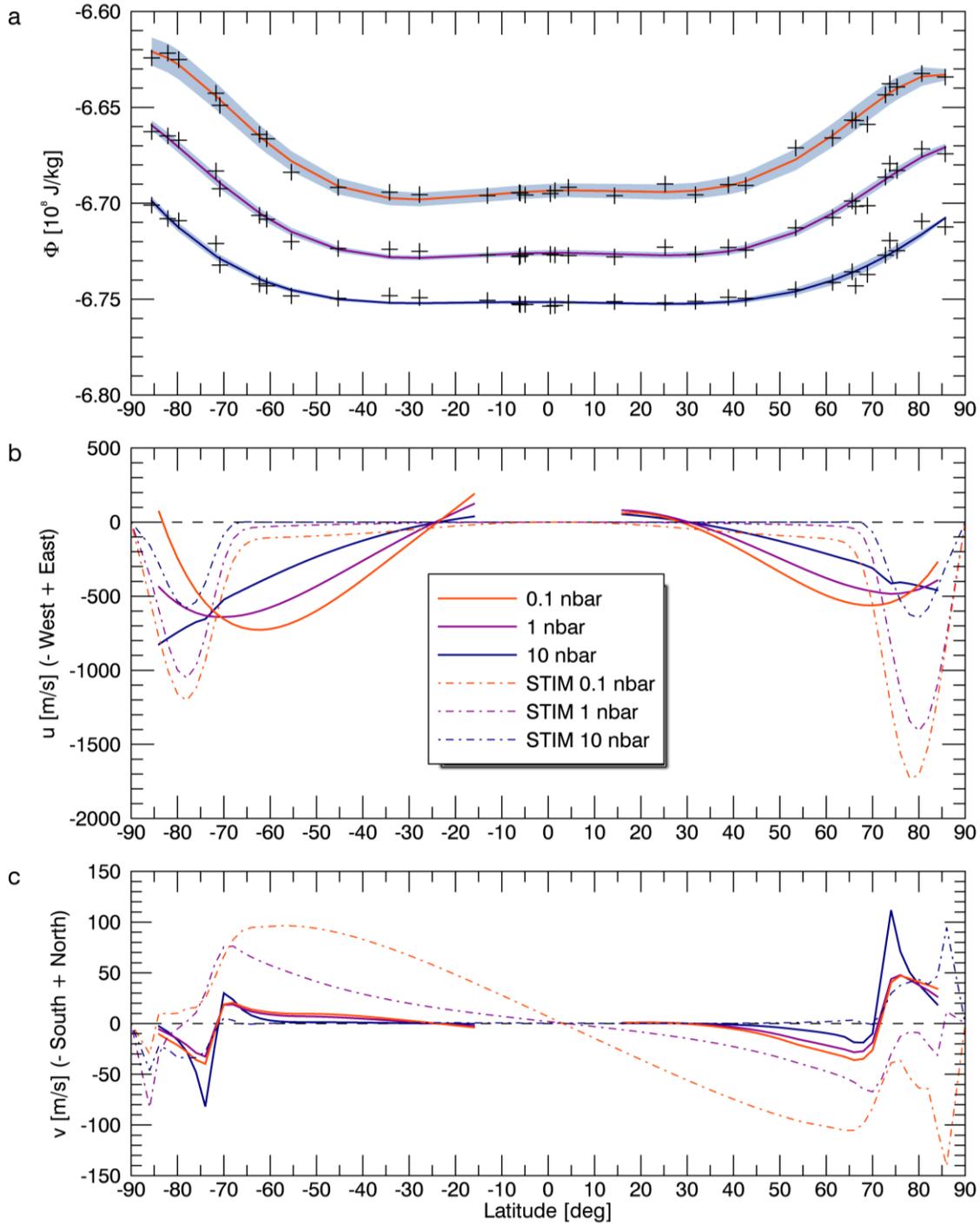

**Fig. 3 | Estimated winds a**, Six-degree polynomial fits to the gravitational potential at 10 nbar (blue), 1 nbar (purple) and 0.1 nbar (red). The shaded region represents the gravitational potential at each latitude plus and minus the 1σ uncertainty in gravitational potential as described in the text. Zonal (**b**) and meridional (**c**) winds derived from the equations of balanced flow (solid lines) and zonally averaged values from the STIM 2019 model[1] (dot dashed lines) at three pressure levels. Our estimated wind profiles are limited to < 86° by the highest latitude observation, and to > 15°, below which the assumption of geostrophy breaks down.

The inferred zonal winds exhibit broad, westward jets spanning from 30° to the highest latitudes sampled and centered at 60-75° latitude at the 1 and 0.1 nbar levels (Fig. 3b). Maximum wind speeds for each level are between 500 and 800 m/s. These jets are driven almost entirely by heating and Coriolis forces. The effect of ion drag can be seen in the small westward enhancements at the 10 nbar level near 75° latitude, corresponding to the sharp peak in auroral electric field and conductivities. Zonal jets predicted by the STIM 2019 model are ~ 10° wide, peak near 80° latitude and exhibit maximum speeds of ~1,100 m/s in the south and ~1,400 m/s in the north (Fig. 3b). In contrast, the inferred zonal winds are broader, slower and extend to lower latitudes than in the model.

Elevated equatorial potential on each of the sampled pressure levels is consistent with eastward winds at latitudes lower than about 30° (Fig. 3b). Winds below this region in the equatorial stratosphere are characterized by Saturn's quasi-periodic oscillation (QPO): a pair of descending and opposing zonal jets[16] imposed upon a strong background eastward jet[17]. The stratospheric jet persists at least up to the 10 mbar level[18] and our data suggest that it is also present in the thermosphere. We note that our gravitational potential does not take into account the effect of differential rotation which was recently detected in Saturn's deep interior [19] and is characterized by super- and sub-rotating jets that extend thousands of kilometers below the 1 bar level. While incorporating the refined gravity coefficients that arise from differential rotation could affect the morphology of the elevated equatorial potential, we estimate that it would produce only small changes to the overall wind field that we infer.

The accuracy of our inferred winds depends on the assumption of modified geostrophy, input parameters such as conductivities and electric fields, and the latitude resolution of our data.

Neglecting the effect of the model input parameters on derived winds, the uncertainty in zonal geostrophic winds depends on the uncertainty in the gradient of the gravitational potential, which is set by our polynomial fit to the data (Fig. 3a). The uncertainty in the potential itself is small, and is derived from uncertainty in pressure level radii. We calculate pressure with the ideal gas law and propagate the error in number density and temperature to find the uncertainty in pressure[20]. The change in radius corresponding to the uncertainty in pressure is given by the hypsometric equation. We bracket the gravitational potential at each latitude by the potential given by the radius at the isobar plus and minus the uncertainty in radius. The mean uncertainty in radius at the 1 nbar pressure level is 20 km, giving an uncertainty in the potential of less than 0.1%. The latitude spacing of our data, however, limits the resolution of the fits and while strong, narrow jets may be present, they are not necessarily detected in the data. Despite our inability to fully resolve the wind structure, it is clear that the jets extend to a lower latitude than in the model and that this arises from the structure of the meridional temperature gradient.

The implied meridional winds diverge near 70° latitude, with flows toward the poles and toward the equator down to a latitude of ~30° (Fig. 3c). Maximum poleward wind speeds are 30-45 m/s at the 0.1 and 1 nbar levels and 85-110 m/s at the 10 nbar level. The peak equatorward flow speed is 20-40 m/s, depending on the pressure level. We expect the thermosphere to be fed from below at latitudes where meridional winds diverge to support conservation of mass. Flow away from 70° latitude suggests the transport of heat to both higher latitudes and to the middle latitudes. This latitude is poleward of the highest observed temperatures and equatorward of 75° where average auroral brightness peaks[21].

In our inferred wind calculations, we neglect viscous and wave drag, which we expect to play a role in circulation. Our assumption of zonal symmetry means that meridional geostrophic winds are zero and therefore only ion drag is available to drive meridional winds. In reality, zonal asymmetries are likely and it is reasonable to expect that we have underestimated meridional wind speeds. A meridional wind pattern similar to the data is predicted by the STIM 2019 model (Fig. 3c), although the winds in the model are in general faster and diverge at a higher latitude than we infer. This indicates that the STIM 2019 model is correct in principle and the drag and heating rates could be adjusted to fit the new data. This adds to the growing evidence contradicting model predictions that Saturn's strong Coriolis forces and polar ion drag prevent equatorward flow of energy[21].

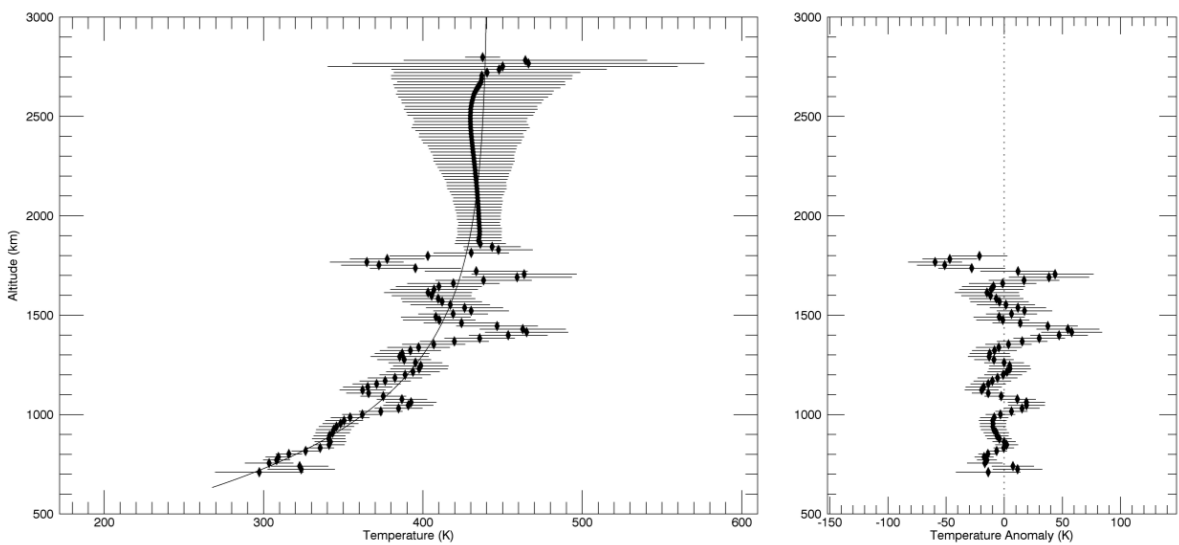

**Fig. 4 | Temperature Profile**. The direct temperature inversion (scatter points) and forward model temperature profile (line) for occultation ST17D202L86N at 86° N latitude (left). Error bars are derived from a Monte Carlo fit to the density profile[4]. The retrieval forces the atmosphere to be isothermal above the altitude of 1,700-1,800 km where the density uncertainty is large. Difference in direct inversion and forward model temperatures below at altitudes below 1,800 km (right). Similar deviations in atmospheric temperature are observed in all 23 Grand Finale observations and may represent atmospheric waves. Waves at higher altitudes are more likely to be amplified by, or artifacts of, the inversion processes.

Müller-Wodarg et al.[1] applied different drag parameterizations to match the previously observed temperatures and we use the best fit, case C (see Fig. 2 of that work). They argue that drag could be from momentum deposition by atmospheric waves and conclude that density perturbations observed by the Cassini Ion Neutral Mass Spectrometer (INMS) measurements during Grand Finale deep dips are sufficiently large to account for the required drag. We observe similar wave-like features in temperature in each of the occultations that we analyzed (Fig. 4) with wavelengths of ~100-200 km and amplitudes on the order of 10-50 K at lower altitudes that in some cases increase to 100 K or more near 1,800 km. Signal to noise in transmission, however, decreases rapidly with altitude and wave properties at high altitudes are uncertain. More work is needed to assess how well wave properties can be inferred from occultation data in general, since these observations probe line of sight densities, but in principle we confirm the detection of waves in the thermosphere.

The pole-to-pole snapshot of Saturn's thermosphere we created from the Cassini Grand Finale occultation data has several implications for circulation and redistribution of energy, including the polar temperature minima that imply less efficient poleward transport of energy or polar upwelling that is not matched by current models. At low latitudes, inferred zonal winds arising from the gravitational potential gradient along isobars are eastward, echoing observed stratospheric flow. Poleward of 30° latitude, we infer slower westward zonal jets that extend to latitudes lower than predicted by models. Our data also support the idea that waves are present globally in the thermosphere and have properties similar to those that can significantly alter global circulation[2]. The results of this study provide evidence for the redistribution of auroral energy to lower latitudes and underline the importance of dynamics in controlling thermospheric temperatures.

# Methods

In general, the Grand Finale occultations are of good quality, obtained with stable pointing and an average altitude sampling resolution of 11 km with the 8.875 s integration time. In this article we refer to occultations by their year, day of year and latitude as designated in Supplementary Tables 1 and 2. Our labels are more concise than those used by the Planetary Data System (PDS), which are also listed.

**Data Evaluation and Selection**

Out of 30 Grand Finale occultations, we analyzed 23 and discarded or postponed 7 due to issues with pointing or insufficient data in critical altitude regions. We describe here our methods for ensuring accurate data retrieval and the circumstances leading to each of the excluded observations. We first inspected detector images and spectra at each time step for anomalies, including evidence of cosmic ray impacts. One in over 11,000 detector images inspected showed evidence of such an anomaly, with all other images demonstrating expected behavior (greatest signal on the center of the detector and small changes between subsequent time steps). Next we checked for stable spacecraft pointing during the observations. This is important because detector response is sensitive to the location of the stellar image in the field of view and drifts in pointing cause time-dependent deviations in observed wavelengths and signal. We found that the majority of Grand Finale occultations are stable. A few of them have periodic drifts of less than 0.06 mrad in the instrument field of view (ST17D182L78S, ST17D185L39N, ST17D198L66N, ST17D202L80S), which are nevertheless acceptably small. We searched the spectra above the altitude of 1,000 km for shifts in wavelength greater than a few picometers that could prevent accurate $H_2$ density retrieval. We identified two occultations with significant wavelength shifts throughout (ST17D178L65S, ST17D201L40S) and decided to postpone to their analysis for future work. In six occultations we found large pointing-induced shifts in wavelength that were limited to high altitudes (ST17D194L34S, ST17D195L55S, ST17D195L45S, ST17D201L28S, ST17D202L80S, ST17D209L61S). We analyzed these occultations, excluding the regions with poor stability, which do not affect the altitudes of interest.

We excluded two occultations (ST17D182L78S, and ST17D196L81N) that were split into sections, probably during downlink. In both occultations, one section covers lower altitudes including the retrieval region and the other section covers high altitudes including the reference region. The observations in each section show slight differences in pointing and further work is needed before they can be analyzed. One occultation is missing data between the altitudes of 1,351 and 1,768 km, in the critical retrieval region, and cannot be analyzed (ST17D189L50N). Two other observations lack sufficient high altitude data to provide a robust reference spectrum, with upper altitude limits of 3,121 and 3,350 km (ST17D176L12N, ST17D176L24N). These observations could be analyzed by using a reference spectrum from another occultation of the same star, but this results in yet greater difficulties due to pointing differences and we exclude them from this work. Finally, because of the varying response of the EUV detector to the stable position of the stellar image in the field of view, variations in absolute wavelength also occur between occultations. We determined the wavelength scale for each occultation by cross-correlating the observed transmission spectra with a model spectrum of the $H_2$ Lyman and Werner absorption bands at altitudes from 1,000 to 1,800 km and then apply this wavelength correction to the other altitudes.

**Data Reduction**

For each occultation, we generate line of sight transmission spectra at individual tangent altitudes by dividing the transmitted spectrum by the unattenuated reference spectrum of the star. Tangent altitudes are defined as the distance along the surface normal to the line of sight from the 1 bar level, which we determine using the Anderson and Schubert[23] model. We measure absorption in the Lyman and Werner electronic bands of $H_2$ between 910 and 1,085 Å. Although the EUV channel observes wavelengths from 580 – 1,180 Å, starlight at wavelengths shorter than 911 Å is almost entirely absorbed by interstellar atomic hydrogen. Because diffusive separation causes the density of species heavier than $H_2$ to drop off rapidly above the homopause located at 0.01 to 0.1 μbar[24], absorption in the EUV channel is dominated by $H_2$. Our analysis excludes the lowest altitudes available in the EUV channel where $CH_4$ begins to contribute to absorption.

In order to calculate transmission, we designate the following altitude regions for each occultation: 1) the reference region, taken at high altitudes where the observed starlight is unocculted 2) the retrieval region, where attenuation in the atmosphere results in transmission between 0 and 1 and 3) the background region, where starlight is fully occulted and any observed photons are from non-stellar sources. The reference regions have lower boundaries between 3,500 km - 5,000 km and upper boundaries between 4,000 km – 11,000 km, depending on the occultation. In designating reference regions, we verify that transmission equals unity in the strong $H_2$ absorption band, 960-970 Å, and exclude any regions of unstable pointing. Because methane also absorbs at EUV wavelengths where we measure $H_2$ absorption, we wish to exclude altitudes where absorption by $CH_4$ is significant from the retrieval regions. Hydrocarbon retrievals will be presented in future work. To minimize absorption by $CH_4$, we define the lower limit of the retrieval regions as the higher of a) the lowest altitude where the running mean over 5 points in the 960-970 Å band exceeds 2 counts and b) the lowest altitude with significant transmission at 1,140-1,150 Å, where $CH_4$ is the main absorber. Finally, we designate the background count region to the altitudes below which both the 960 – 970 Å band and the 1,140-1,150 Å methane band transmission light curves go to zero (from 450 to 750 km depending on the occultation). Within this region we determine the background counts for each wavelength, taken as the average over the altitudes in the background region. We find small average background counts of 0 to 10 per wavelength per integration period, which we subtract from the signal at all altitudes.

After these checks, we retrieve the $H_2$ density and temperature profiles from the transmission time series by using the methods adapted from Koskinen et al.[4], which we summarize here. The column density at each tangent altitude is a measure of the total atmospheric $H_2$ along the line of sight. We retrieve column density using a Levenberg–Marquardt least squares algorithm to fit the observed transmission spectra with a high resolution transmission model based on the $H_2$ line lists of Abgrall et al.[25,26,27] convolved with the line spread function of the UVIS instrument (see UVIS User's Guide available through the PDS Rings Node), generalized to the EUV channel. Since the absorption cross section of $H_2$ in the Lyman and Werner electronic bands depends on temperature (which we are working to derive), we retrieve the column density iteratively, starting with an initial $H_2$ cross section at a constant temperature of 300 K.

We invert the column density profiles by using Tikhonov regularization and fit a forward model temperature profile[20] to the resulting number density profile. We then re-retrieve column densities, now using $H_2$ cross sections that vary with altitude according to the forward model temperatures from the previous step. This is repeated until subsequent solutions converge (between 4 and 10 iterations). We then invert column density again with Tikhonov regularization to obtain the final number density profile. Assuming hydrostatic equilibrium and an ideal gas, we also integrate the number density profiles to directly retrieve temperature profiles which typically agree well with the parameterized forward model temperature profile. The number density profiles above the altitude of 1,700-1,800 km are generally noisy and we smooth the profiles in this region by fitting the data with an exponential power series. Although the relatively large uncertainties prevent us from ruling out high altitude temperature gradients, the fits are generally consistent with an isothermal atmosphere. Because we expect temperatures to remain constant with altitude out to the exobase at 2,700 – 3,000 km[28], we have adopted the common convention of referring to the temperature above 1,700-1,800 km as the "exospheric" temperature. Our results indicate a trend in exothermic temperature with latitude in the auroral regions. Given that auroral brightness is not uniform in local solar time[29], we considered the possibility that our latitudinal trend in polar exospheric temperatures might be correlated instead with auroral activity. We find this trend is not primarily dependent upon local solar time (see Fig. S2).

Figure 4 shows an example of the direct inversion and forward model temperature profiles for an occultation at 86 N. We detect wave-like features in temperature in all of the 23 Grand Finale occultations, which constitutes evidence for upwardly propagating gravity waves. Because the average temperature profile resolution is ~20 km in altitude, only waves with wavelengths a few times this distance are detectable. Furthermore, integration along the line of sight means we can only detect waves with a large horizontal extent. More work is needed to interpret the observations and accurately infer the properties of the detectable waves.

**Derivation of Winds**

We infer zonal wind speeds at three pressure levels based on equations 16 and 17 from Larsen and Walterscheid[3] used for the terrestrial northern hemisphere thermosphere. Based on conservation of momentum, these equations assume a balance of Coriolis, pressure gradient, and ion drag forces, which are the three terms represented within the brackets below:

$$\frac{d\vec{v}}{dt} = \frac{1}{\rho}\left[-2\rho\vec{\Omega} \times \vec{v} - \nabla P + \vec{J} \times \vec{B}\right] = 0 \qquad (1)$$

Here v is the horizontal wind vector, $\rho$ is the neutral mass density, $\Omega$ is Saturn's angular velocity, $\nabla P$ is the pressure gradient, J is the current density and B is the magnetic field.

Under the assumption of zonal symmetry in isobaric coordinates, the pressure gradient force is represented by the gradient of the gravitational potential with respect to latitude on surfaces of constant pressure that we determine from the forward model $H_2$ densities and temperatures. For most occultations, the latitude varies by less than one degree over the retrieval region and we take latitude at the half-light point to be constant. To find the gravitational potential, we interpolate radius to surfaces of constant pressure at 10, 1 and 0.1 nbar based on the retrieved density and temperature profiles. We calculate gravitational potentials by using zonal harmonics ($J_{2n}$) and rotation rate (10:32:35h ± 13s) from Anderson and Schubert[23]. While new results provide updated zonal harmonics[30], rotation rate (10:33:34h ± 55s) and reveal differential rotation[19], we estimate that only minor changes to our inferred winds would result from these new values. For consistency with the STIM 2019 model and our previous work, we use the Anderson and Schubert values. We then fit a six-degree polynomial to the gravitational potentials and estimate zonal geostrophic winds from the potential gradient.

We use spheroidal latitudes to determine the gravitational potential gradient with respect to latitude[12], $\frac{\partial \Phi}{\partial \varphi}$, which allows us to determine the geostrophic component of the winds:

$$u_g = \frac{-1}{f(\varphi)R} \frac{\partial \Phi(p,\varphi)}{\partial \varphi} \quad (2)$$

$$v_g = 0 \quad (3)$$

Here $u_g$ is the zonal geostrophic wind and $v_g$ is the meridional geostrophic wind, which we take to be zero since our temperature profile is zonally agnostic. R is the equatorial radius of Saturn taken to be 60,268 km and $f = 2\Omega\sin(\varphi)$ is the latitude-dependent Coriolis parameter. We take $\Omega$ to be 1.6554302 x $10^{-4}$ radians/s based on the model of Anderson and Schubert[23].

Following Larsen and Walterscheid, we define the Lorentz force in terms of the Hall and Pedersen conductivities, $\sigma_h$ and $\sigma_p$, the electric field, E, neutral velocity, $v_n$, and radial magnetic field, $B_r$:

$$\vec{J} \times \vec{B} = \begin{bmatrix} \sigma_p(p,\varphi) & \sigma_h(p,\varphi) & 0 \\ -\sigma_h(p,\varphi) & \sigma_p(p,\varphi) & 0 \\ 0 & 0 & \sigma_0 \end{bmatrix} \cdot [\vec{E}(\varphi) + \vec{v_n}(p,\varphi) \times \vec{B_r}(p,\varphi)] \times \vec{B_r}(p,\varphi) \quad (4)$$

We use electric field values from Jia et al. (2012)[11], that depend only on latitude and are the same as those used by the STIM 2019 model. The electric field is based on field-aligned currents that connect the ionosphere to the magnetosphere and are driven by mass loading in the magnetosphere and solar wind interaction, as explained in Cowley et al. (2004b)[10] and Jia et al. (2012)[11]. Zonally averaged Pedersen and Hall conductivities are calculated by STIM and depend on both pressure and latitude. Radial magnetic field values depend on pressure and latitude and are based on the Saturn Pioneer Voyager (SPV) model of Saturn's magnetic field from Davis and Smith[31]. While updated magnetic field parameters have been recently provided by Dougherty et al. [8], these do not significantly alter the derived wind speeds and we use previous values for consistency with the STIM 2019 model. The assumption that the magnetic field is radial is best at the poles and breaks down at low latitudes. We do not include ion drag at middle and low latitudes that can be driven by wind-driven electrodynamics but is assumed not to be important here. Zonal and meridional plasma drift velocities, $u_p$ and $v_p$ respectively, are:

$$u_p = \frac{-E_{south}}{B_r(p,\varphi)} \tag{5}$$

$$v_p = \frac{-E_{east}}{B_r(p,\varphi)} \tag{6}$$

Solving simultaneously the equations zonal and meridional force balance we derive the following equations for zonal winds, u (positive eastward), and meridional winds, v (positive northward), similar to equations 16 and 17 in Larsen and Walterscheid (1995)[3]:

$$u = \frac{(1-\beta b_r)u_g + [\alpha^2 - \beta b_r(1-\beta b_r)]u_p + \alpha(v_p - v_g)}{\alpha^2 + (1-\beta b_r)^2} \tag{7}$$

$$v = \frac{(1+\beta b_r)v_g + [\alpha^2 - \beta b_r(1-\beta b_r)]v_p + \alpha(u_p - u_g)}{\alpha^2 + (1-\beta b_r)^2} \tag{8}$$

where α and β depend on the Pedersen and Hall terms respectively and are defined as:

$$\alpha = \frac{\sigma_p(p,\varphi)B_r^2(\varphi)}{f(\varphi)\rho(p,\varphi)} \tag{9}$$

$$\beta = \frac{\sigma_h(p,\varphi)B_r^2(\varphi)}{f(\varphi)\rho(p,\varphi)} \tag{10}$$

We have implemented the factor, $b_r$, to account for the difference between Larsen and Walterscheid's terrestrial northern hemisphere formulation and Saturn's magnetic field direction over both hemispheres. This term is +1 in the northern hemisphere and -1 in the southern hemisphere. We note that ion drag is implicit and included in the equations 16 and 17 of Larsen and Walterscheid[3]. The magnitude of this term depends on the magnetospheric electric field that is based on field-aligned currents coupling to the magnetosphere[11] and conductivities in the ionosphere that are taken from the STIM 2019 model. In these balanced flow equations, the role of polar ion drag is surprisingly limited and the steady state circulation depends mostly on the observed thermal structure, presumably driven by polar Joule heating and redistribution of energy by dynamics.

## Acknowledgements
Z.B. and T.T.K. acknowledge support from the NASA Cassini Data Analysis Program (80NSSC19K0902). R.W.'s contributions were carried out at the Jet Propulsion Laboratory, under a contract with NASA (80NM0018D0004). L.W.E. was supported by the Cassini Project.

## Author Contributions
Z.L.B. performed the analysis, developed analytical tools and wrote the manuscript. T.T.K. developed analytical tools, conceived of the strategy to infer horizontal winds and provided feedback on the manuscript. I.M.W. provided guidance on the interpretation of the results, STIM outputs, conductivity and electric field data. R.W., A. J. and T.T.K designed the observing campaign and commented on the manuscript. L.W.E. is the PI of the UVIS instrument who contributed to UVIS observation campaigns and commented on the manuscript.

## Data Availability

The data that support the plots within this paper and other findings of this study are available from the NASA Planetary Data System (https://pds-imaging.jpl.nasa.gov/portal/cassini_mission.html) or from the corresponding author upon reasonable request

**Correspondence and requests for materials** should be addressed to Z.B.

# Supplementary Information

| ID | PDS | Date | Star | Lat [°] | Lon [°] | LST [hrs] | Vertical Resolution [km] | Temp [K] |
|---|---|---|---|---|---|---|---|---|
| ST16D014L25N | EUV2016_014_21_05 | 1/14/2016 | ε Ori | 25.2 | -19.4 | 19.2 | 32.7 | 470 ± 42 |
| ST16D045L04N | EUV2016_045_02_15 | 2/14/2016 | α Vir | 4.4 | 76.5 | 2.6 | 48.9 | 411 ± 17 |
| ST16D045L01N | EUV2016_045_08_50 | 2/14/2016 | α Vir | 1.5 | 50.7 | 14.8 | 0.0 | 361 ± 15 |
| ST16D046L66N | EUV2016_046_23_30 | 2/15/2016 | γ Ori | 65.6 | 29.0 | 4.7 | 32.4 | 526 ± 34 |
| ST16D094L00N | EUV2016_094_18_01 | 4/3/2016 | γ Ori | 0.5 | 23.1 | 18.7 | 53.2 | 371 ± 27 |
| ST16D296L14N | EUV2016_296_06_40 | 10/22/2016 | ζ Cen | 14.3 | 44.7 | 16.3 | 43.8 | 399 ± 35 |
| ST16D339L06S | EUV2016_339_04_47 | 12/4/2016 | β Cru | -6.2 | -54.5 | 2.2 | 83.3 | 401 ± 17 |
| ST16D353L06S | EUV2016_353_13_13 | 12/18/2016 | β Cru | -6.0 | -171.4 | 2.1 | 47.2 | 403 ± 16 |
| ST17D009L13S | EUV2017_009_00_55 | 1/9/2017 | α Cru | -13.1 | 67.3 | 3.5 | 24.5 | 391 ± 14 |
| ST17D087L05S | EUV2017_087_21_53 | 3/28/2017 | β Cru | -5.0 | 173.6 | 1.7 | 48.4 | 385 ± 19 |
| ST17D175L71S | EUV2017_175_19_31 | 6/24/2017 | ε Ori | -70.8 | -111.5 | 6.7 | 28.0 | 478 ± 16 |
| ST17D175L82S | EUV2017_175_20_36 | 6/24/2017 | ζ Ori | -82.1 | -166.6 | 7.5 | 26.6 | 450 ± 8 |
| ST17D182L86S | EUV2017_182_06_15 | 7/1/2017 | ε Ori | -85.6 | 157.6 | 12.9 | 27.8 | 373 ± 13 |
| ST17D182L32N | EUV2017_182_20_40 | 7/1/2017 | ε Ori | 31.8 | -64.9 | 5.7 | 41.1 | 340 ± 31 |
| ST17D183L43N | EUV2017_183_00_47 | 7/2/2017 | ζ Ori | 42.6 | 159.1 | 5.7 | 36.3 | 478 ± 18 |
| ST17D185L39N | EUV2017_185_17_28 | 7/4/2017 | β CMa | 38.9 | 156.0 | 7.5 | 31.2 | 516 ± 20 |
| ST17D188L72S | EUV2017_188_17_43 | 7/7/2017 | ε Ori | -71.7 | 5.4 | 17.0 | 30.6 | 452 ± 15 |
| ST17D188L62S | EUV2017_188_20_24 | 7/7/2017 | ζ Ori | -62.3 | -85.0 | 17.5 | 30.2 | 510 ± 14 |
| ST17D192L53N | EUV2017_192_06_25 | 7/11/2017 | β CMa | 53.4 | -91.3 | 8.1 | 30.4 | 547 ± 17 |
| ST17D194L34S | EUV2017_194_14_52 | 7/13/2017 | γ Ori | -34.2 | -78.5 | 17.0 | 39.9 | 465 ± 24 |
| ST17D195L74N | EUV2017_195_02_00 | 7/14/2017 | γ Ori | 73.7 | -16.8 | 21.4 | 45.1 | 476 ± 11 |
| ST17D195L55S | EUV2017_195_05_17 | 7/14/2017 | ε Ori | -55.4 | 156.2 | 17.4 | 41.0 | 478 ± 36 |
| ST17D195L45S | EUV2017_195_08_35 | 7/14/2017 | ζ Ori | -45.3 | 47.8 | 17.7 | 33.2 | 487 ± 23 |
| ST17D195L69N | EUV2017_195_21_44 | 7/14/2017 | ε Ori | 68.9 | 151.6 | 5.1 | 31.0 | 499 ± 19 |
| ST17D198L66N | EUV2017_198_18_21 | 7/17/2017 | β CMa | 66.4 | 63.1 | 9.5 | 41.0 | 586 ± 22 |
| ST17D201L28S | EUV2017_201_21_41 | 7/20/2017 | ζ Ori | -27.8 | 143.8 | 17.8 | 31.5 | 394 ± 15 |
| ST17D202L86N | EUV2017_202_09_04 | 7/21/2017 | ε Ori | 85.7 | -127.7 | 0.3 | 37.0 | 437 ± 11 |
| ST17D202L75N | EUV2017_202_11_31 | 7/21/2017 | ζ Ori | 75.3 | 63.6 | 18.7 | 35.4 | 454 ± 10 |
| ST17D202L80S | EUV2017_202_21_18 | 7/21/2017 | κ Ori | -79.8 | 107.2 | 20.5 | 27.3 | 439 ± 16 |
| ST17D205L73N | EUV2017_205_05_21 | 7/24/2017 | β CMa | 72.8 | -92.4 | 12.6 | 38.1 | 455 ± 14 |
| ST17D209L61S | EUV2017_209_09_02 | 7/28/2017 | κ Ori | -60.7 | -142.3 | 18.7 | 37.6 | 513 ± 21 |
| ST17D210L61N | EUV2017_210_04_22 | 7/29/2017 | κ Ori | 61.4 | 121.6 | 6.7 | 31.0 | 523 ± 24 |
| ST17D216L80N | EUV2017_216_16_10 | 8/4/2017 | κ Ori | 80.6 | -74.0 | 8.7 | 32.7 | 457 ± 16 |

**Supplementary Table 1 | Occultation Data Analyzed.** UVIS stellar occultation parameters used in this work for the Grand Finale and earlier in 2017 and 2016. Planetocentric latitude and longitude and local solar time are taken at the half light point. Exospheric temperatures fit to altitudes above 1,700-1,800 km are listed.

| ID | PDS | Date | Star | Lat [°] | Lon [°] | LST [hrs] |
|---|---|---|---|---|---|---|
| ST17D176L12N | EUV2017_176_06_40 | 6/25/2017 | ε Ori | 12.3 | -125.8 | 5.8 |
| ST17D176L24N | EUV2017_176_11_17 | 6/25/2017 | ζ Ori | 24.0 | 81.9 | 5.8 |
| ST17D178L65S | EUV2017_178_14_32 | 6/27/2017 | β CMa | -65.2 | 101.6 | 3.7 |
| ST17D182L78S | EUV2017_182_08_43 | 7/1/2017 | ζ Ori | -78.2 | 136.1 | 16.9 |
| ST17D182L78S | EUV2017_182_09_19 | 7/1/2017 | ζ Ori | -78.2 | 136.1 | 16.9 |
| ST17D189L50N | EUV2017_189_09_54 | 7/8/2017 | ε Ori | 49.7 | 23.8 | 5.5 |
| ST17D189L50N | EUV2017_189_10_07 | 7/8/2017 | ε Ori | 49.8 | 22.7 | 5.5 |
| ST17D189L50N | EUV2017_189_10_09 | 7/8/2017 | ε Ori | 49.8 | 21.6 | 5.5 |
| ST17D196L81N | EUV2017_196_00_54 | 7/15/2017 | ζ Ori | 80.5 | 24.5 | 4.6 |
| ST17D196L81N | EUV2017_196_01_34 | 7/15/2017 | ζ Ori | 80.4 | 23.1 | 4.6 |
| ST17D201L40S | EUV2017_201_17_42 | 7/20/2017 | ε Ori | -39.6 | -85.5 | 17.6 |

**Supplementary Table 2 | Occultation Data Not Analyzed.** UVIS Grand Finale stellar occultations not analyzed in this work. Planetocentric latitude and longitude and local solar time are taken at the half light point. Exospheric temperatures fit to altitudes above 1,700-1,800 km are listed.

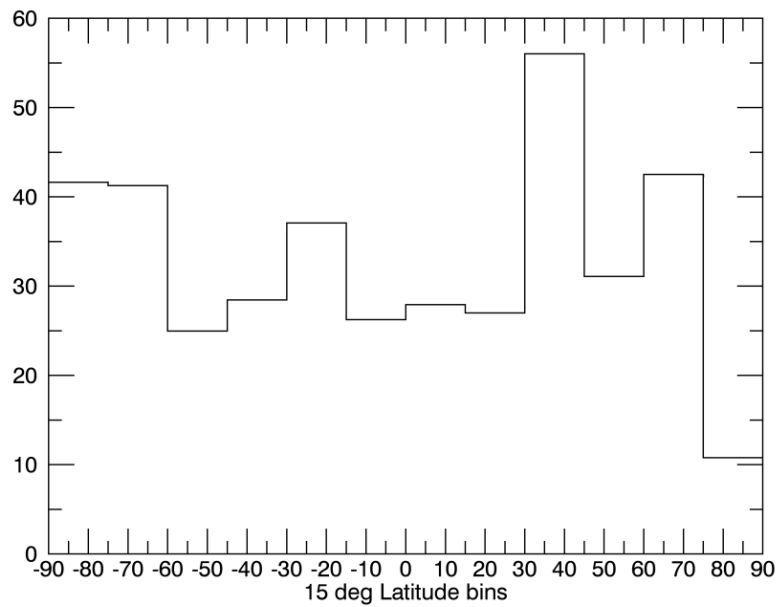

**Supplementary Figure 1 | Temperature Variability.** Standard deviation in temperature within 15° latitude bins for UVIS stellar and solar occultations taken 2004-2017 from Koskinen et al. (2013) and Koskinen et al. (2015). The greatest variability exists in the 30-45° N latitude bin.

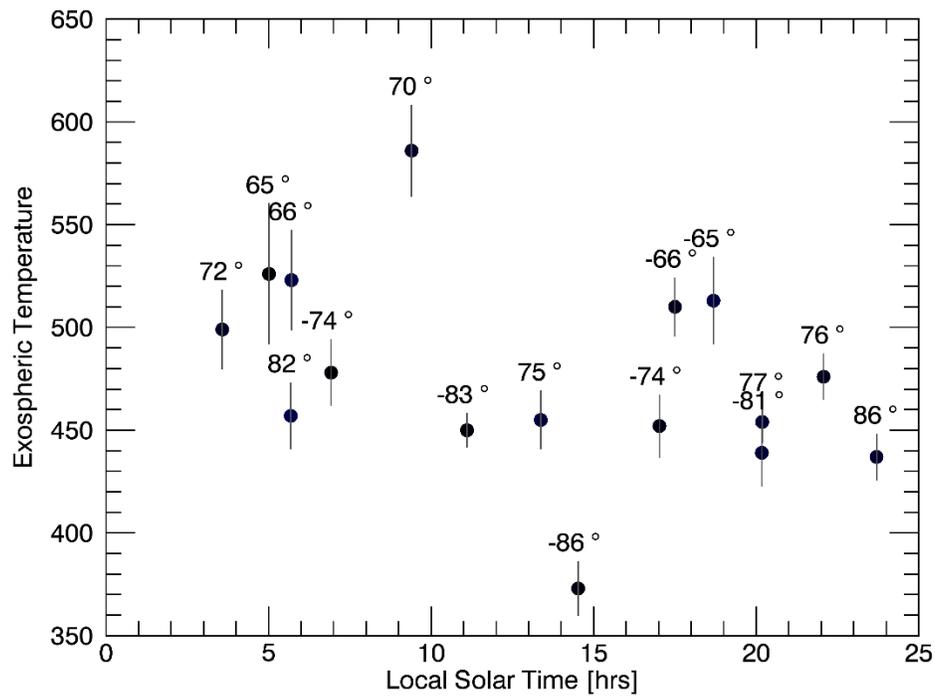

**Supplementary Figure 2 | High Latitude Temperatures with Local Time.** Exospheric temperature as a function of local time for Grand Finale UVIS observations for observations made poleward of 65° latitude. Temperature does not appear to be correlated with local time.